\documentstyle[11pt,newpasp,twoside,epsf]{article}
\index{summary}
\index{instructions}
\index{template}

\def\edcomment#1{\iffalse\marginpar{\raggedright\sl#1\/}\else\relax\fi}
\reversemarginpar

\begin{document}
\title{The BeppoSAX View on the 2001 Reactivation of SGR 1900+14}

\author{M. Feroci $^{1}$, S. Mereghetti $^{2}$, E. Costa$^{1}$,
J.J.M. in 't Zand$^{3}$, P. Soffitta$^{1}$, T. Cline$^{4}$, R. Duncan$^{5}$, 
M. Finger$^{6}$, S.V. Golenetskii$^{7}$,
K. Hurley$^{8}$, C. Kouveliotou$^{6}$, P. Li$^{8}$, E. Mazets$^{7}$,
M. Tavani$^{2}$, C. Thompson$^{9}$, P. Woods$^{6}$}
\affil{
$^{1}$ {\it Istituto di Astrofisica Spaziale - CNR, Rome, Italy}\\
$^{2}$ {\it Istituto di Fisica Cosmica - CNR, Milan, Italy}\\
$^{3}$ {\it SRON and Astron. Inst., Utrecht University, Utrecht, 
            The Netherlands}\\
$^{4}$ {\it NASA/GSFC, Greenbelt, MD, USA}\\
$^{5}$ {\it Univ. of Texas, Department of Physics, Austin, TX, USA}\\
$^{6}$ {\it NASA/MSFC, Huntsville, AL, USA}\\
$^{7}$ {\it Ioffe Physico-Technical Institute, S. Petersburg, Russia}\\
$^{8}$ {\it Space Science Laboratory, Univ. of California, Berkeley, CA, USA}\\
$^{9}$ {\it Canadian Institute of Theoretical Physics, Toronto, Canada}
}

\begin{abstract}

After a couple of years of quiescence, the soft gamma repeater
SGR 1900+14 suddenly reactivated on 18 April 2001, with the emission
of a very intense, long and modulated flare, only second in
intensity and duration to the 27 August 1998 giant flare.
BeppoSAX caught the large flare with its Gamma Ray Burst Monitor 
and with one of the Wide Field Cameras.
The Wide Field Cameras also detected
X-ray bursting activity shortly before the giant flare.
A target of opportunity observation was started
only 8 hours after the large flare with the Narrow Field Instruments,
composed of two 60-ks long pointings.
These two observations show an X-ray afterglow of the persistent
SGR 1900+14 source, decaying with time according to a power law of
index -0.6.

\end{abstract}

\section{The 2001 reactivation of SGR 1900+14}

On April 18, 2001 at 07:55:11.509 UT the Gamma Ray Burst Monitor
(GRBM, 40-700 keV) onboard BeppoSAX was triggered by a large flare.
The event was also observed by the unit 1 of the BeppoSAX
Wide Field Cameras (WFCs, 2-28 keV), that was automatically shut-down
few seconds after the start of the event by an onboard safeguard algorithm.
The WFC localized this giant flare as originating from
the soft gamma-ray repeater SGR 1900+14.

The BeppoSAX GRBM unit 1, co-aligned to WFC unit 1, 
detected a peak flux of 16,400 counts/s (40-700 keV, 
after deadtime correction), corresponding to (1.1$\pm$0.1)$\times10^{-5}$
erg cm$^{-2}$ s$^{-1}$. The event lasted approximately 40~s,
and the total fluence was about 240,000 counts.
A 5-s periodic modulation of the light curve is consistent with
the 5.17-s period of the persistent X-ray source.
In the light curve of the flare we can identify the
repetition of 5 or 6 cycles, and a last pulse that is out of
phase with respect to the previous ones.
Assuming a thermal bremsstrahlung spectral shape, from our 2-channel, 
1-s resolved spectrum, we can derive a plasma temperature kT of $\sim$30~keV.
There is no evidence for a hard spike at the start of the event, and
the spectral evolution during the event is relatively smooth, with a general
hardening trend.

The Konus-Wind instrument detected this event in the non-trigger
mode because of very high background level after strong solar flare
(30 times more than normal). Our estimation for energy  more than
15 keV is: fluence - 2$\times$10$^{-4}$ erg cm$^{-2}$ and peak flux -
2.5$\times$10$^{-5}$ erg cm$^{-2}$ s$^{-1}$. 
In spite of rough time resolution in
the background mode the Konus-Wind data has evidences of the 5-s
pulsations. 

Three weak and short bursts are detected at times
2537, 755 and 444 seconds {\it before} the giant flare in the WFC unit 1. 
Their durations
are 100, 125 and 55 ms, and their peak fluxes exceed 20 Crab units
in 2-28 keV (Feroci et al. 2001b). No gamma-ray counterpart was 
found for these events in the GRBM, nor in {\it Ulysses} or Konus/Wind.
This is the first reported bursting activity from SGR 1900+14 since
approximately two years.

At the time of the giant flare, as detected by the GRBM, the WFC1
detected a large increase in the count rate for approximately 3
seconds after the GRBM trigger. At this time the count rate was
so high (larger than 30,000 counts/s) 
that it triggered the automatic turn off of the experiment.

\section{BeppoSAX NFI observations}

The April 18 2001 activity of SGR 1900+14 allowed us to activate our
Target of Opportunity observation program with the BeppoSAX Narrow
Field Instruments (NFI, 0.1-300 keV).
Thanks to the remarkable effort by the
Mission Scientist Team, the Mission Planners Team and our team,
the first follow-up
observation was started on April 18 at 15:10 UT, less than 8 hours
after the event. This first observation (Feroci et al. 2001c) 
was 28 hours long, for a
net exposure time in the MECS instrument of 60 ks. A second pointing 
was carried out on April 29 at 20:20 UT, for a similar exposure time.
The persistent X-ray counterpart of SGR 1900+14 was detected in
both observations. The source emitted a large number
of short X-ray bursts during the first observation, whereas
in the second pointing only one burst was detected.

The average spectrum during the first pointing, after removal
of the bursts, can be fit with a power law with photon index
2.6$\pm$0.1 and N$_{H}$=(4.3$\pm$0.3)$\times10^{22}$ cm$^{-2}$.
The persistent 2-10 keV flux showed a rapid decaying behaviour
during our first pointing, going approximately
from 3.5 to 2$\times10^{-11}$ erg cm$^{-2}$ s$^{-1}$ (unabsorbed).
Using the X-ray flux in the same band provided by the Chandra
observation carried out on April 22 (Kouveliotou et al. 2001),
we obtain that the time behaviour
of the X-ray flux can be described by a power law, with index
$\sim$-0.6. Interestingly, studying the hardness ratio between the
1.6-4 keV and the 4-10 keV energy ranges, 
we find a clear correlation flux vs. hardness,
indicating that the spectrum softens when the persistent flux
decreases.

During our second observation the source appeared
to have gone back to its quiescent status. Removing the single burst
that we observed, the 0.1-10 keV energy spectrum is best fit by
the sum of an absorbed power law plus a blackbody with temperature
$\sim$0.5 keV. The 2-10 keV flux is consistent with the `standard'
quiescent status, at $\sim$1.1$\times10^{-11}$ erg cm$^{-2}$ s$^{-1}$
(unabsorbed).
In Figure 1 we show the plot of the X-ray flux vs. time, including
also the Chandra point, that suggests that the source returned to
its quiescent status in less than 3 days after the burst.

\begin{figure}
\epsfxsize=9cm
\centerline{\epsffile{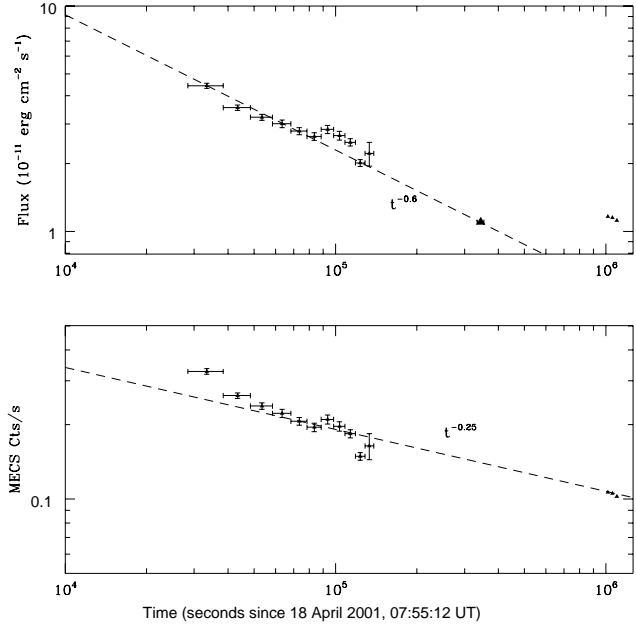}}
\caption{
Time history of the 2-10 keV flux from the persistent
SGR 1900+14 source.
{\it Top}: The two BeppoSAX/NFI observations are shown, as well as one Chandra
observation (Kouveliotou et al. 2001) around t=3.4$\times10^{5}$ s.
{\it Bottom}: only the two BeppoSAX observations are shown.
}
\end{figure}

We performed a period search in both NFI observations and found
the expected 5.17-s periodicity in the 2-10 keV flux from the persistent
source. In the first observation the best
period was 5.17277 s, with a 
smooth sinusoidal profile. In the second pointing the period increased
to 5.17298 s, without significant differences in the pulse profile.
Therefore, the period derivative
between the two BeppoSAX observation is about 2.3$\times10^{-10}$ s s$^{-1}$.

\section{Discussion}

The BeppoSAX instrumentation observed the most interesting phases of
the 2001 activation of SGR 1900+14 with a very complete set of observations.
The GRBM and WFC caught the second giant flare in the (observed) history of
this source and its precursor X-ray activity. The
NFI observed the afterglow of the persistent source after the flare, the
bursting activity and its quenching, and the period history soon after
the giant flare.

The giant flare is rather unusual with respect to its two predecessors, the
March 5 1979 event from SGR 0526-66 and the August 27 1998 flare from
SGR 1900+14 (e.g., Mazets et al. 1979, Hurley et al. 1999, 
Feroci et al. 2001a).
The most important differences in the April 18 are in the following 
aspects:
\begin{itemize}
\item the peak flux and fluence are more than one order of
magnitude smaller
\item the duration is almost one order of magnitude shorter
\item there is no evidence for the first short and very hard pulse at the
beginning of the event, implying that, contrary to the other two
events, most likely there was no relativistic particle emission during
this event.
\end{itemize}

The NFI observation is the fastest pointing after a giant flare and
provides a {\it direct} observation of a long-duration
afterglow emission from an SGR after a giant flare. 
Our result of an afterglow showing a power law decay, with index -0.6, is 
similar to that obtained by Woods et al. (2001), who measured a decay 
index of -0.713$\pm$0.025 after the August 27 event, based on the assumed 
constancy of the pulsed fraction. Interestingly, Ibrahim et al. (2001) found
a similar decay law in the pulsating soft tail of the August 29 short burst.

The spectral variability of the X-ray afterglow is also interesting.
During our first observation, at high flux level, the source spectrum
showed a softening, and an overall power law shape. In our second
observation, at a flux 
typical of the quiescent state,
we clearly detect the emergence of a kT=0.5 keV blackbody component,
and an N$_{H}$ smaller by about a factor of 2.
Based on our observations, together with our analysis of the previous
BeppoSAX observations (see also Woods et al. 1999, 2001) of this source 
in quiescence (May 1997 and March/April
2000) and after the August 27 giant flare (September 1998), we can
state that the blackbody component for this source is peculiar of its 
quiescent state.

\acknowledgments

We warmly thank L. Piro, G. Gandolfi, L. Salotti, M. Capalbi, L. Amati,
the BeppoSAX Mission Planners, the BeppoSAX SOC, OCC and SDC 
for their personal efforts that allowed a fast repointing
of the source after the flare, as well as a prompt analysis of the GRBM, 
WFC and NFI data.

\end{document}